\documentclass[showkeys,notitlepage,superscriptaddress,floatfix,prd,10pt,aps]{revtex4-1}

\usepackage[utf8]{inputenc}
\usepackage[colorlinks=true,citecolor=blue,linkcolor=blue]{hyperref}
\usepackage[normalem]{ulem}
\usepackage{url}
\usepackage{graphicx,wrapfig,float,slashed,cancel}
\usepackage{amsmath,amssymb,epsfig,graphicx,xcolor,stmaryrd}
\usepackage{bbold}
\usepackage{bm}
\usepackage{enumitem}
\usepackage{multirow}

\usepackage{physics}
\usepackage{orcidlink}

\definecolor{darkblue}{RGB}{1, 90, 173}

\def\pslash{p\!\!\!\slash }

\def\qslash{q\!\!\!\slash }
\def\xslash{x\!\!\!\slash }

\def\dslash{D\!\!\!\slash }

\begin{document}


\title{Mechanical properties of proton using flavor-decomposed gravitational form factors}
\author{Zeinab Dehghan\orcidlink{0000-0002-5976-9231}}
\email{zeinab.dehghan@ut.ac.ir}
\thanks{Corresponding author}
\affiliation{Department of Physics, University of Tehran, North Karegar Avenue, Tehran 14395-547, Iran}
\author{F. Almaksusi\orcidlink{0009-0002-5932-7967}}
\email{fateme.almaksusi@gmail.com}
\affiliation{Department of Physics, University of Tehran, North Karegar Avenue, Tehran 14395-547, Iran}
\author{K. Azizi\orcidlink{0000-0003-3741-2167}}
\email{kazem.azizi@ut.ac.ir}
\thanks{Corresponding author}
\affiliation{Department of Physics, University of Tehran, North Karegar Avenue, Tehran
	14395-547, Iran}
\affiliation{Department of Physics, Do\v{g}u\c{s} University, Dudullu-\"{U}mraniye, 34775
	Istanbul,  T\"{u}rkiye}

\date{\today}

\preprint{}

\begin{abstract}

We investigate the mechanical properties of the proton by extracting its flavor-decomposed gravitational form factors (GFFs) using Light-Cone QCD sum rules (LCSR). These form factors encode critical information about the internal dynamics and spatial distribution of energy, momentum, and internal forces within the proton. The flavor decomposition of the quark sector indicates the role of each flavor in  the proton’s pressure and shear force distributions. Our results show that the up quark contributes more significantly compared to the down quark in the three conserved proton GFFs, as well as in the energy and shear force distributions. Additionally, we define the non-conserved form factor $\bar{c}^q (t)$, which takes part in the distributions of energy and pressure; the latter is essential for maintaining proton stability. Furthermore, we determine the proton's mass and mechanical radii, providing valuable insight into its internal structure and dynamics.

\end{abstract}



\maketitle

\renewcommand{\thefootnote}{\#\arabic{footnote}}
\setcounter{footnote}{0}
\section{Introduction}\label{intro}

Gravitational form factors (GFFs) provide fundamental insight into the internal structure of hadrons by characterizing their interaction with the energy-momentum tensor (EMT)~\cite{ Kobzarev:1962wt, Pagels:1966zza}. Unlike electromagnetic form factors that examine charge and magnetic distributions, GFFs reveal crucial information about the mass, spin, pressure and shear force distributions within hadrons, playing a key role in understanding their mechanical structure~\cite{Burkert:2023wzr}. Although GFFs have been explored in various hadronic systems~\cite{Pefkou:2021fni,Alharazin:2022wjj,Fu:2022bpf,Dehghan:2023ytx,Fu:2023ijy,Goharipour:2024atx,He:2023ogg, Wang:2023bjp,Hackett:2023nkr}, the nucleon remains a primary focus of investigation. Despite its early discovery, the proton's internal structure is not yet fully understood, and its gravitational form factors are the most significant tools for probing its internal dynamics and mechanical properties.

Nucleon GFFs have been investigated through various methods and theoretical frameworks, making their study a vibrant and evolving field in both experimental and theoretical physics. The first parametrization of GFFs for spin-1/2 systems was introduced in the pioneering works of~\cite{Kobzarev:1962wt, Pagels:1966zza}. Using the conserved QCD energy-momentum tensor, the nucleon’s gravitational structure is characterized by three fundamental form factors: $A$, $J$, and $D$, which describe the mass,
total angular momentum, and the spatial distribution of internal forces, respectively~\cite{Ji:1996ek, Wang:2021vqy, Liu:2021lke}. However, when decomposing the EMT into its quark and gluon contributions, an additional form factor, $\bar{c}$, emerges. This form factor contributes to the quark and gluon pressure distributions inside the proton and governs the internal force balance between quark and gluon subsystems, crucial for maintaining proton stability~\cite{Polyakov:2018exb}. The $\bar{c}$ form factor is associated with the trace anomaly in QCD and highlights the distinct role of gluons in the nucleon’s mass structure~\cite{Liu:2021gco, Tanaka:2022wrr, Tanaka:2022wzy, Liu:2023cse, Wang:2024lrm}. Furthermore, the non-conserved $\bar{c}$ form factor, potentially relating to the cosmological constant as a metric term in Einstein’s equations, has been speculated to be connected to the vacuum energy contribution in general relativity~\cite{Liu:2021gco, Liu:2023cse}.  Investigating this form factor could provide deeper insights into the role of the trace anomaly in hadron mass generation and its possible implications for fundamental questions in quantum field theory and cosmology.

Empirical measurement of the proton gravitational form factors requires studying graviton-proton scattering. However, since gravity is negligible at small scales like proton size, direct measurement of GFFs is not currently feasible. Instead, indirect measurements of these form factors were proposed through Generalized Parton Distribution Functions (GPDs) in Refs.~\cite{Collins:1996fb,Ji:1996ek,Ji:1996nm,Radyushkin:1997ki}. Several studies have employed the GPD approach with different experimental data to extract the proton GFFs and their properties~\cite{Kumericki:2019ddg,Burkert:2018bqq,Selyugin:2023hqu,Wang:2024sqg,Dutrieux:2021nlz,Wang:2024fjt,Goharipour:2025lep,Goharipour:2024mbk,Hashamipour:2022noy,Mamo:2019mka,Bhattacharya:2023ays,Won:2023ial,Shuryak:2023siq,Chen:2024adg,Bhattacharya:2023wvy, Freese:2021czn, Freese:2021qtb} (for reviews on various GPD methods, see e.g.~\cite{Anikin:2017fwu,Burkert:2023wzr}).  In recent decades, the gravitational form factors of the nucleon have gained much more attention and have been explored through various physical frameworks. These include lattice QCD, the light-front quark-diquark model, the holographic QCD model, chiral effective field theory (EFT), the Skyrme model, and others~\cite{Bali:2018zgl,Alexandrou:2019ali,Lin:2023ass, GarciaMartin-Caro:2023klo, Kou:2023azd, Alharazin:2023uhr, GarciaMartin-Caro:2023toa,Yao:2024ixu,Cao:2023ohj,Panteleeva:2022uii,Panteleeva:2021iip,Panteleeva:2020ejw,Won:2022cyy,Hatta:2023fqc,Choudhary:2022den,Neubelt:2019sou,Tong:2022zax,Cao:2024fto,Hagiwara:2024wqz}. Studies have also investigated the gluonic component of the QCD energy-momentum tensor current of the proton in different approaches~\cite{Meziani:2024cke, More:2023pcy, Guo:2023pqw, Guo:2023qgu, Duran:2022xag, Fan:2022qve, Mamo:2021krl}. In Refs.~\cite{More:2023vlb, Wang:2023fmx} the gluon and quark components have been examined individually. Additionally, Ref.~\cite{Burkert:2021ith} presents the first determination of the shear forces on quarks inside the proton, derived from experimental data of Deeply Virtual Compton Scattering (DVCS). The quark contribution to the EMT current has also been studied using light-cone QCD sum rules and transverse-momentum distributions (TMDs) in Refs.~\cite{Lorce:2023zzg, Anikin:2019kwi, Azizi:2019ytx}. The proton’s mechanical properties, such as internal pressure and shear forces, have been investigated through various theoretical approaches, with gravitational form factors playing a central role~\cite{Chakrabarti:2020kdc, Sain:2025kup, Sugimoto:2025btn, Fujii:2025aip, Nair:2024fit, Goharipour:2025yxm, Goharipour:2025lep, Hackett:2023rif, Won:2023zmf}.

One particularly intriguing aspect of proton gravitational form factors is their decomposition into individual flavor contributions. This decomposition of flavors is essential for understanding how quarks and gluons contribute to the proton’s mechanical structure, enabling a more precise characterization of the momentum, pressure, and shear force distributions within the proton. Recent theoretical and lattice QCD studies have provided significant insights into the behavior of flavor-decomposed GFFs and the spatial distribution of internal forces, uncovering nontrivial effects such as the distinct mechanical roles of valence and sea quarks, as well as the necessity of gluon contributions~\cite{Nair:2024fit, Hackett:2023rif, Won:2023cyd, Amor-Quiroz:2023rke, Won:2023zmf, Freese:2020mcx, Alexandrou:2017oeh}. Some studies on flavor decomposition of GFFs consider both quark and gluon components, while others focus exclusively on the quark sector. In the quark sector of the proton's energy-momentum tensor current, as well as in its decomposition into individual flavors, the system is not strictly conserved. As a result, the non-conserved form factor $\bar{c}$ appears, providing a means to investigate the effects of non-conservation in the EMT on the proton's internal properties, such as the distribution of internal forces and the mechanisms contributing to its stability in greater detail. A recent lattice QCD study~\cite{Hackett:2023rif} defines proton GFFs by considering the full EMT current and performing a flavor decomposition that focuses on the contributions from both valence quarks (up and down) and the sea quark (strange) to study the mechanical properties of the proton. The flavor structure of EMT focusing on $\bar{c}(t)$ form factor is presented in Ref.~\cite{Won:2023cyd}. The flavor decomposition of proton gravitational form factors has been derived using a basis light-front quantization approach in Ref.~\cite{Nair:2024fit}.

The study of proton form factors provides another crucial concept which is its size. The proton size is commonly characterized by its charge radius~\cite{Xiong:2019umf}, which describes the spatial distribution of charge within the proton and is determined from electromagnetic form factors. While other radii such as mass and mechanical are derived from gravitational form factors. The $D(t)$ form factor provides valuable information about the structure and shape of nucleons. The mechanical radius quantifies the distribution of internal forces within the proton, while the mass radius reflects the spatial distribution of energy associated with the quark and gluon fields. Notably, the nucleon’s charge radius is larger than both its mechanical and mass radii. This is expected, as electromagnetic interactions extend over longer distances, whereas the strong interaction is more
compact, concentrating more toward the proton’s core. Recent studies have utilized experimental data, such as meson photoproduction, to extract the proton mass radius and other mechanical properties~\cite{Wang:2022uch,Wang:2022zwz,Mamo:2022eui,Cao:2024zlf,Kharzeev:2021qkd}. The analysis~\cite{Kharzeev:2021qkd} suggests that the root mean square (rms) mass radius of the proton is approximately $0.55$ fm, which is significantly smaller than the rms charge radius of about $0.84$ fm. This difference highlights the distinct nature of the electromagnetic and strong interactions.

In the present study, the GFFs of the valence quark components of the proton energy-momentum tensor are calculated employing the light-cone QCD sum rules (LCSR) method. The nucleon GFFs have been studied with this approach without regard to flavor decomposition~\cite{Anikin:2019kwi, Azizi:2019ytx}.
We utilize Distribution Amplitudes (DAs), which encapsulate both perturbative (soft) contributions and non-perturbative effects, as they are expanded across different twists. DAs emerge in hadronic matrix elements, where the products of quark current operators are expanded near the light cone.
The energy-momentum tensor describes the interaction between gravitational-like and matter fields, in our specific case, the nucleon fields, similar to how gauge fields interact with particles via the corresponding electromagnetic current~\cite{Braun:2001tj}. We employ the DAs of the nucleon as derived in~\cite{Braun:2000kw, Braun:2006hz} to determine the flavor-decomposed gravitational form factors of the proton. The capability of the LCSR method to predict the physics lying behind the hadrons makes it a complementary technique to lattice QCD in the calculation of hadron form factors~\cite{Khodjamirian:2020btr}.

As mentioned, in this paper, we focus on the quark contribution to the EMT current of the proton. This component can be further decomposed into singlet and triplet currents of the valence flavors. We extract GFFs of each flavor and conduct a comprehensive investigation of their mechanical properties. The rest of the paper is structured as follows. In Sec.~\ref{sec:flavordicompos}, we introduce the flavor-decomposed EMT and the corresponding proton form factors. The GFFs of the proton, calculated using the light-cone sum rules framework, are presented in Sec.~\ref{sec:lightcone}. In Sec.~\ref{sec:numresults}, we provide a numerical analysis of the proton GFFs for individual valence quarks and their flavor combinations. In Sec.~\ref{sec:mech properties}, we use these gravitational form factors to define the distributions of energy, shear force, and pressure for different proton flavors. Additionally, we examine key mechanical properties of the proton, including its internal structure, mass, and mechanical radii. Finally, in Sec.~\ref{sec:conclusion}, we conclude our work with a discussion of the obtained results.

\section{Flavor decomposition of energy-momentum tensor}\label{sec:flavordicompos}

The energy-momentum tensor (EMT) current is,
\begin{equation}
T_{\mu\nu}^{q+g} = \sum_q T_{\mu\nu}^q + T_{\mu\nu}^g ,
\label{eq:totalEMT}
\end{equation}
which includes contributions from both quark and gluon components. The explicit forms of the quark and gluon EMT currents are given by~\cite{Polyakov:2018zvc},
\begin{eqnarray}
\label{eq:EMTcurrentquark}
T_{\mu\nu}^q (x) &=& \frac{i}{4} \bar{q} (x) \bigg(\overleftrightarrow{D}_\mu \gamma_\nu
+ \overleftrightarrow{D}_\nu \gamma_\mu \bigg) q(x)
- g_{\mu\nu} \bar{q}(x)  \big(\frac{i}{2} \overleftrightarrow{\dslash} - m_q\big)q(x),
\\
T_{\mu\nu}^g (x) &=& G_{\mu\rho}(x) G^{\rho}_{\hspace{0.5mm},\nu}(x) + \frac{1}{4}g_{\mu\nu}G^{\rho\delta}(x)G_{\rho\delta}(x).\label{eq:EMTcurrentgluon}
\end{eqnarray}
For the derivation of energy-momentum tensor of gauge theories see Ref.~\cite{Freese:2021jqs}. The total EMT current is conserved, satisfying ${\partial}^{\mu} T_{\mu\nu} = 0$. The covariant derivative $\overleftrightarrow{D}_{\mu}$ is expressed as $\overleftrightarrow{D}_\mu (x) = \overrightarrow{D}_\mu (x) - \overleftarrow{D}_\mu (x)$ with,
\begin{equation}\label{eq:derivdetail}
\overrightarrow{D}_{\mu}(x)=\overrightarrow{\partial}_{\mu}(x)-i g A_\mu(x), \qquad
\overleftarrow{D}_{\mu}(x)=\overleftarrow{\partial}_{\mu}(x)+
i g A_\mu(x),
\end{equation}
and $A_\mu(x)$ is the external gluon field. In the quark sector, the flavor decomposition of the quark EMT current for proton is described as follows,
\begin{equation}\label{eq:singletEMT}
\hspace{-1cm}\text{singlet:} \quad T_{\mu\nu}^q = T_{\mu\nu}^{u+d} = T_{\mu\nu}^u + T_{\mu\nu}^d, \qquad (I = 0),
\end{equation}
\begin{equation}\label{eq:tripletEMT}
\text{triplet:} \quad T_{\mu\nu}^u, \quad T_{\mu\nu}^{u-d} = T_{\mu\nu}^u - T_{\mu\nu}^d, \quad T_{\mu\nu}^d, \qquad (I = 1).
\end{equation}
In these expressions, $I$ denotes the isospin, with $u+d$ and $u-d$ representing the isoscalar and isovector components of the EMT current, respectively. In this study, we separately extract the gravitational form factors (GFFs) using the singlet and triplet quark currents for the proton system.

The matrix element of the EMT current between proton states can be expressed as~\cite{Polyakov:2018zvc},
\begin{align}\label{eq:matrix element}
\begin{aligned}
\langle {N(p',s')}
|T_{\mu \nu}^j(x)|
{N(p,s)}\rangle &= 
\bar{u}(p',s')\Big\{ \frac{P_{\mu}P_{\nu}}{m} A^j(t) 
+ \frac{i}{2} \frac{(P_{\mu} \sigma_{\nu\rho} + P_{\nu} \sigma_{\mu\rho})
\Delta^{\rho}}{m} J^j(t) 
+ \frac{(\Delta_{\mu}\Delta_{\nu} - g_{\mu\nu} \Delta^2)}{4 m} D^j(t) \\
&\hspace{2cm}+ m g_{\mu\nu} \bar{c}^j(t) \Big\} u(p,s) e^{i(p' - p).x}, 
\end{aligned}
\end{align}
where $j = q, g, q+g$ and $u(p,s)$ is the Dirac spinor with momentum $p$ and spin $s$, $P = (p + p')/2$, $\Delta = p' - p$, $t = \Delta^2$, $\sigma_{\mu\rho}=\frac{i}{2}[\gamma_{\mu},\gamma_{\rho}]$ and $m$ is the mass of proton. The form factors $A^j(t)$ and $J^j(t)$ are called the mass and angular momentum form factors, respectively. The $D^j(t)$ form factor provides information about the internal mechanical structure of the proton, including pressure and shear force distributions. The term $\bar{c}^j(t)$ is a non-conserved form factor, which vanishes for a conserved EMT current:$\sum_{q}\bar{c}^q(t) + \bar{c}^g(t) = 0$. Thus, for a conserved EMT, only three form factors remain.

The $\bar{c}^{q,g}(t)$ form factors emerge when considering the non-conserved EMT currents and contribute to the quark and gluon pressure distributions inside the proton, thus we highlight its importance. The cosmological constant term $(\Lambda)$ in Einstein’s equation of general relativity is as follows~\cite{Einstein:1917ce}:
\begin{equation}
R_{\mu\nu} - \frac{1}{2} R g_{\mu\nu} + \Lambda g_{\mu\nu} = 8 \pi G ~T_{\mu\nu},
\label{einstein}
\end{equation}
where $R_{\mu\nu} $ is the Ricci curvature tensor, $R$ is scalar curvature, $g_{\mu\nu}$ is the metric tensor, $T_{\mu\nu}$ is the stress-energy tensor and $G$ is Newton's constant. 
When the vacuum energy is included in Einstein's equations, it can be manifested as the cosmological constant.
In the matrix element of the QCD EMT in Eq.~\eqref{eq:matrix element}, the $\bar{c}(t)$ form factor is the coefficient of the metric tensor $g_{\mu\nu}$, which is alike the cosmological constant term in Eq.~\eqref{einstein}. 
On the other hand, the $\bar{c}(t)$ form factor corresponds to the trace part of the energy-momentum tensor, directly tied to the trace anomaly.
The trace anomaly contributes to the vacuum expectation value of EMT and consequently to the cosmological constant. The trace anomaly in QCD arises from both quark and gluon contributions to the energy-momentum tensor. 
The gluon part of the trace anomaly is particularly significant, as it accounts for a substantial portion of the nucleon mass. Recent studies have focused on isolating the glue part of the trace anomaly form factors for hadrons like the pion and nucleon, providing deeper insights into the role of gluonic contributions in mass generation~\cite{Wang:2024lrm, Hu:2024mas, Liu:2023cse}. 

In the following section, we utilize QCD sum rules to derive four gravitational form factors for each flavor of the proton.

\section{QCD sum rules}\label{sec:lightcone}

We employ the light-cone sum rules (LCSR) to calculate the flavor-decomposed gavitational form factors of the proton, using the following two-point correlation function,
\begin{equation}
\Pi_{\mu\nu}^j(p,q) = i \int d^4 x e^{-iq.x}
\langle 0 |\mathcal{T}[J_{N}(0) T_{\mu \nu}^j(x)]| N(p) \rangle,
\label{eq:corrf}
\end{equation}
where $\mathcal{T}$ denotes the time ordering operator, $p$ ($p'$) is the four-momentum of the initial (final) proton, $q = p' - p$ is the momentum transfer, and $J_{N}$ is  the interpolating current for the nucleon. We determine proton GFFs in different flavors, $j = u, d, u+d, u-d$, while disregarding gluon fields. The derivation of gluon field contributions in LCSR requires the quark-gluon mixed distribution amplitudes of the nucleon, which are currently unavailable. In this work, we consider the chiral limit: $m_u = m_d = 0$. In Eq.~\eqref{eq:corrf}, the general form for the nucleon interpolating current is represented as,
\begin{equation}
J_{N}(x) = 2 \varepsilon_{abc} \Big[ \big( u^{aT}(x) C d^{b}(x) \big) \gamma_{5} u^{c}(x) 
+ \beta  \big( u^{aT}(x) C \gamma_{5} d^{b}(x) \big) u^{c}(x)   \Big],\label{eq:ninterpolating}
\end{equation}
with $C$ denoting the charge conjugation operator; and $a, b$ and $c$ are color indices, and $\beta$ is an arbitrary mixing parameter.

In the framework of QCD sum rules, we evaluate the correlation function of different quark flavors from two sides: the physical (hadronic) side and the QCD side. In the following sections, we will provide a detailed derivation of the correlation function pertaining to the quark sector of the energy-momentum tensor.

\subsection{Physical side of the correlation  function}\label{subsec:physical side}

First, we concentrate on determining the hadronic side of the flavor-decomposed correlation function.
To this end, we insert a complete set of intermediate nucleon states~\cite{Khodjamirian:2020btr, Colangelo:2000dp} into Eq.~\eqref{eq:corrf}, 
\begin{equation} \label{eq:CompeletSet}
1=\vert 0\rangle\langle0\vert +\sum_{h}\int\frac{d^4 p'_h}{(2\pi)^4}(2\pi) \delta(p'^2_h-m^2)|h(p'_h)\rangle  \langle h(p'_h)|+\mbox{higher Fock states},
\end{equation}
where the sum is over all states with the same content and quantum numbers as the nucleon and with momentum $p'_h$. We then use the following identity,
\begin{equation} 
\label{eq:dxdp}
\int d^4 x\int\frac{d^4 p'_h}{(2\pi)^4}(2\pi) \delta(p'^2_h-m^2)
e^{i(p'_h - p').x}
= \frac{i}{m^2 - p'^2},
\end{equation}
which helps us 
isolate the ground-state nucleon pole contribution after some calculations. As a result, the hadronic side of the correlation function takes the form~\cite{Azizi:2019ytx},
\begin{align}\label{eq:phys}
\Pi_{\mu\nu}^{Had, j} (p,q) =& \displaystyle\sum_{s{'}} \frac{\langle0|J_N(0)|{N(p',s')}\rangle\langle {N(p',s')}
	|T_{\mu \nu}^j(0)|N(p,s)\rangle}{m^2-p'^2} +\cdots,
\end{align}
where dots contribute to the continuum, excited and multi-hadron states.
The second matrix element in the numerator of Eq.~\eqref{eq:phys} is defined in terms of the gravitational form factors in Eq.~\eqref{eq:matrix element}. However, the first matrix element is written in terms of the nucleon's residue  $\lambda_{N}$~\cite{Ioffe:1981kw, Aliev:2002ra, Aliev:2011ku, Azizi:2014yea},
\begin{equation}
\label{eq:rezi}
\langle0|J_N (0)|{N(p',s')} \rangle = \lambda_{N} u (p',s').
\end{equation}
The residue, as a spectroscopic parameter, plays a crucial role in calculating various physical observables related to the interaction and decay properties of hadrons. In this work, we utilize the nucleon residue to derive an analytical expression for the physical side of the correlation function and eventually compute the gravitational form factors. 

By substituting the matrix elements from Eqs.~\eqref{eq:matrix element} and~\eqref{eq:rezi} into the correlation function in Eq.~\eqref{eq:phys}, and then applying the following completeness relation:
\begin{equation}
\label{eq:complete}
\sum_{s'} u(p', s') \bar{u}(p', s') = \pslash' + m,
\end{equation}
we derive the hadronic side of the correlation function expressed in terms of the proton's gravitational form factors, 
\begin{widetext}
\begin{align}
\label{eq:had}
\Pi_{\mu\nu}^{Had, j}(p,q)=&\frac{\lambda_N }{{m^2-p'^2}} (\pslash' + m)
\Big\{ \frac{P_{\mu}P_{\nu}}{m} A^j(t) 
+ \frac{i}{2} \frac{(P_{\mu} \sigma_{\nu\rho} + P_{\nu} \sigma_{\mu\rho})
	\Delta^{\rho}}{m} J^j(t) 
+ \frac{(\Delta_{\mu}\Delta_{\nu} - g_{\mu\nu} \Delta^2)}{4 m} D^j(t) \nonumber\\
&\hspace{3cm}+ m g_{\mu\nu} \bar{c}^j(t) \Big\} u(p,s), 
\end{align}
\end{widetext}
After using Dirac equation $\pslash u(p,s) = m u(p,s)$ and then performing the Borel transformations on the variable $p'^2=(p+q)^2$ with Borel parameter $M^2$ in addition to continuum subtraction, we get,
\begin{align}
\label{eq:HadStr}
\Pi_{\mu\nu}^\text{Had, j}(Q^2) &= \lambda_{N} e^{-\frac{m^2}{M^2}} \Big[ 
\Pi_{1}^\text{Had, j} (Q^2) p_{\mu} p_{\nu} \mathbb{1}
+ \Pi_{2}^\text{Had, j} (Q^2) q_{\mu} q_{\nu} \mathbb{1}
+ \Pi_{3}^\text{Had, j} (Q^2) p_{\mu} q_{\nu} \mathbb{1}
+ \Pi_{4}^\text{Had, j} (Q^2) g_{\mu\nu} \mathbb{1} \nonumber \\
&+ \Pi_{5}^\text{Had, j} (Q^2) p_{\mu} p_{\nu} \qslash 
+ \Pi_{6}^\text{Had, j} (Q^2) q_{\mu} q_{\nu} \qslash
+ \Pi_{7}^\text{Had, j} (Q^2) p_{\mu} q_{\nu} \qslash +\cdots\Big].
\end{align}
where $Q^2 = -t$ and $\mathbb{1}$ is the unit matrix. 
We use the $\lambda_N$ expression calculated using QCD two-point sum rule method as below~\cite{Aliev:2002ra, Aliev:2011ku},
\begin{align}
\label{eq:residue}
\lambda_N^2 =~& e^{m_N^2/M^2} \Bigg\{\frac{M^6}{256 \pi^4} E_2(x) (5+2 \beta + \beta^2) 
- \frac{\langle \bar{q}q \rangle^2}{6} \Big[6 (1-\beta^2)  -
(1-\beta)^2  \Big] + \frac{m_0^2}{24 M^2} \langle \bar{q}q \rangle^2 \Big[12 (1-\beta^2) - (1-\beta)^2  \Big]\Bigg\}, 
\end{align}
with $x = s_0/M^2$ where $s_0$ is the continuum threshold and $E_n(x)$ defined as,
\begin{eqnarray}
\label{nolabel}
E_n(x)=1-\sum_{i=0}^{n}e^{-x}\Big(\frac{x^i}{i!}\Big)~. \nonumber
\end{eqnarray}
The functions $\Pi_{i}^\text{Had, j} (Q^2)$ in Eq.~\eqref{eq:HadStr} include the desired flavor-decomposed gravitational form factors of the proton. We present some of the structures for the sake of brevity. In the following we investigate the QCD side of the correlation function.

\subsection{QCD side of the correlation function}\label{subsec:QCD side}

To calculate the QCD side, we substitute Eqs.~\eqref{eq:EMTcurrentquark},~\eqref{eq:tripletEMT}, and~\eqref{eq:ninterpolating} into correlation function~\eqref{eq:corrf}, focusing on the valance quarks, up and down. After applying Wick's theorem to Eq.\eqref{eq:corrf} and evaluating all possible contractions, the QCD correlation functions of up and down flavors are obtained as follows:
\begin{equation}
\Pi_{\mu\nu}^{u}(p,q) = - \frac{1}{2}\int d^4 x e^{-iqx}\Bigg[R 
\Big (\delta^\sigma_\alpha \delta^\theta_\rho \delta^\phi_\beta S(-x)_{\delta \omega}
+\, \delta^\sigma_\delta \delta^\theta_\rho \delta^\phi_\beta S(-x)_{\alpha \omega}\Big)
\langle 0|\varepsilon_{abc} u_{\sigma}^a(0) u_{\theta}^b(x) d_{\phi}^c(0)|N(p)\rangle
+\, (\mu \leftrightarrow \nu) \Bigg],
\label{eq:upcorrfunc}
\end{equation}
and
\begin{equation}
\hspace{-3.7cm}\Pi_{\mu\nu}^{d}(p,q) = - \frac{1}{2}\int d^4 x e^{-iqx}\Bigg[R 
\,\,
\delta^\sigma_\alpha \delta^\theta_\delta \delta^\phi_\rho S(-x)_{\beta \omega} 
\,\langle 0|\epsilon^{abc} u_{\sigma}^a(0) u_{\theta}^b(0) d_{\phi}^c(x)|N(p)\rangle
+\, (\mu \leftrightarrow \nu) \Bigg],
\label{eq:downcorrfunc}
\end{equation}
with $R$ defined as follows,
\begin{equation}
R = \Big(C_{\alpha\beta} (\gamma_5)_{\xi\delta}
+ \beta (C \gamma_5)_{\alpha\beta}\, (\mathbb{1})_{\xi\delta}
\Big)\Big( \overleftrightarrow{D}_\mu \gamma_\nu
- g_{\mu\nu} \overleftrightarrow{\dslash}
\Big)_{\omega \rho} .
\label{eq:Acoeff}
\end{equation}
In Eqs.~\eqref{eq:upcorrfunc} and~\eqref{eq:downcorrfunc}, 
$S(x)$ denotes the light quark propagator, which in the chiral limit, $m_u=m_d=0$, is given by~\cite{Aliev:2008cs},
\begin{align}
	S(x)&= \frac{i\,\xslash}{2\,\pi^2 x^4}-\frac{\langle q\bar{q}\rangle}{12}\left(1+\frac{m_0^2 x^2}{16}\right)
	-ig_s\int^1_0 d\upsilon\left[\frac{\xslash}{16\pi^2 x^4} G_{\mu\nu}\sigma^{\mu\nu}
-  \frac{i\, \upsilon\, x^\mu}{4\pi^2 x^2} G_{\mu\nu}\gamma^\nu\right],
\label{eq:propagator}
\end{align}
where ${\langle q\bar{q}\rangle}$ represents the quark condensate, and $m_0^2 = \langle\bar{q}g_s G^{\mu\nu}\sigma_{\mu\nu}q\rangle / \langle\bar{q}q\rangle$ corresponds to the quark-gluon condensate. In this context, gluon interactions are excluded. Additionally, the quark condensate contributions will be suppressed by the Borel transformation. The up correlation function in Eq.~\eqref{eq:upcorrfunc} includes two propagators due to the valance structure of proton which involves two up quarks. The correlation functions of isocalar and isovector flavor combination are obtained through summing and subtracting the correlation functions of up~\eqref{eq:upcorrfunc} and down~\eqref{eq:downcorrfunc}, respectively. The QCD correlation functions in Eqs.~\eqref{eq:upcorrfunc} and ~\eqref{eq:downcorrfunc} require evaluating the matrix elements of quark operators inserted between the vacuum and nucleon states, expressed as,
\begin{equation}
\langle 0|\varepsilon_{abc} u_{\sigma}^a(a_1 x) u_{\theta}^b(a_2 x) d_{\phi}^c(a_3 x)|N(p)\rangle,
\label{eq:matrix element quark}
\end{equation}
with $a_1$, $a_2$ and $a_3$ representing real parameters. For the up-quark matrix element in Eq.~\eqref{eq:upcorrfunc}, 
$a_1=a_3=0$, while for the down-quark matrix element in Eq.~\eqref{eq:downcorrfunc}, 
$a_1=a_2=0$. We proceed by expanding these matrix elements based on the nucleon distribution amplitudes (DAs), see Appendix~\ref{ap:DAs}, specifically in Eq.~\eqref{eq:das-def}. Furthermore, we apply Eqs.~\eqref{eq:das-def},~\eqref{eq:DAsdirect},~\eqref{eq:integ} for each flavor and carry out the necessary simplifications such as Wick rotation and Schwinger parameterization on the QCD side correlation function. The next step is to perform the Borel transformation to the variable $p'^2$. This is followed by continuum subtraction to suppress contributions from higher-order terms and continuum states. This process is facilitated by specific replacement rules that are presented below,
\begin{align}
\int_0^1 dx \frac{\rho(x)}{(q+xp)^2}
&\rightarrow 
- \int_{x_0}^1
dx
\frac{\rho(x)}{x} e^{-s(x)/M^2},\nonumber	
\\
\int_0^1 dx \frac{\rho(x)}{(q+xp)^4}
&\rightarrow \frac{1}{M^2} \int_{x_0}^1
dx
\frac{\rho(x)}{x^2} e^{-s(x)/M^2}
+\frac{\rho(x_0)}{Q^2+x_0^2 m^2} e^{-s_0/M^2},\nonumber
\\
\int_0^1 dx \frac{\rho(x)}{(q+xp)^6}
&\rightarrow 
-\frac{1}{2M^4}
\int_{x_0}^1
dx
\frac{\rho(x)}{x^3} e^{-s(x)/M^2}
-\frac{1}{2M^2}\frac{\rho(x_0)}{x_0(Q^2+x_0^2m^2)}e^{-s_0/M^2}
\nonumber\\
&+\frac{1}{2}\frac{x_0^2}{Q^2+x_0^2m^2}\bigg[\frac{d}{dx_0}\frac{\rho(x_0)}{x_0(Q^2+x_0^2m^2)}\bigg]e^{-s_0/M^2},\nonumber	
\\
\int_0^1 dx
\frac{p'^2 \rho(x)}{(q+xp)^2}
&\rightarrow 
-\int_{x_0}^1
dx
\frac{\rho(x)}{x} s(x) e^{-s(x)/M^2},\nonumber	
\\
\int_0^1 dx \frac{p'^2\rho(x)}{(q+xp)^4}
&\rightarrow  \int_{x_0}^1
dx
\frac{\rho(x)}{x^2} 
\Big(-1 + \frac{s(x)}{M^2}\Big)
e^{-s(x)/M^2}
+\frac{\rho(x_0)}{Q^2+x_0^2 m^2} s_0 e^{-s_0/M^2},\nonumber
\\
\int_0^1 dx \frac{p'^2 \rho(x)}{(q+xp)^6}
&\rightarrow 
\frac{1}{M^2}
\int_{x_0}^1
dx
\frac{\rho(x)}{x^3} 
\Big(1 - \frac{s(x)}{2 M^2}\Big)
e^{-s(x)/M^2}
+\frac{1}{2}\frac{\rho(x_0)}{x_0(Q^2+x_0^2m^2)}\Big(1 - \frac{s_0}{M^2}\Big)e^{-s_0/M^2}
\nonumber\\
&+\frac{1}{2}\frac{x_0^2}{Q^2+x_0^2m^2}\bigg[\frac{d}{dx_0}\frac{\rho(x_0)}{x_0(Q^2+x_0^2m^2)}\bigg]
s_0 e^{-s_0/M^2},\nonumber	
\\
\int_0^1 dx \frac{p'^4\rho(x)}{(q+xp)^4}
&\rightarrow  \int_{x_0}^1
dx
\frac{\rho(x)}{x^2} 
\Big(-2 s(x) + \frac{s^2(x)}{M^2}\Big)
e^{-s(x)/M^2}
+\frac{\rho(x_0)}{Q^2+x_0^2 m^2} s^2_0 e^{-s_0/M^2},\nonumber
\\
\int_0^1 dx \frac{p'^4 \rho(x)}{(q+xp)^6}
&\rightarrow 
\int_{x_0}^1
dx
\frac{\rho(x)}{x^3} 
\Big(-1 + \frac{2 s(x)}{M^2} - \frac{s^2(x)}{2 M^4}\Big)
e^{-s(x)/M^2}
+\frac{\rho(x_0)}{x_0(Q^2+x_0^2m^2)} \Big(s_0 -\frac{s^2_0}{2M^2} \Big)e^{-s_0/M^2}
\nonumber\\
&+\frac{1}{2}\frac{x_0^2}{Q^2+x_0^2m^2}\bigg[\frac{d}{dx_0}\frac{\rho(x_0)}{x_0(Q^2+x_0^2m^2)}\bigg]
s^2_0 e^{-s_0/M^2},
\label{eq:BorelSubtract}
\end{align}
with $(q+xp)^{2N} = (-x)^N (s-p'^2)^N$ where,
\begin{eqnarray}
s(x) = \frac{(1-x)}{x} Q^2 + m^2 (1-x),
\end{eqnarray}
and $x_0$ is the solution of the quadratic equation of continuum threshold $s=s_0$:
\begin{eqnarray}
x_0&=&\Big[\sqrt{(Q^2 + s_0 - m^2)^2 + 4m^2 Q^2} - (Q^2 + s_0 - m^2)\Big]/2m^2.
\end{eqnarray}
In this work, considering all the possible Borel transformations  and flavor decomposition,  we derive proton's GFFs more accurately.  After the Borel transformations and continuum subtraction, we obtain an expression for each specific flavor of QCD side correlation function,
\begin{align}
\label{eq:QCDStr}
\Pi_{\mu\nu}^\text{QCD, j}(Q^2) &= \Big[ 
\Pi_{1}^\text{QCD, j} (Q^2) p_{\mu} p_{\nu} \mathbb{1}
+ \Pi_{2}^\text{QCD, j} (Q^2) q_{\mu} q_{\nu} \mathbb{1}
+ \Pi_{3}^\text{QCD, j} (Q^2) p_{\mu} q_{\nu} \mathbb{1}
+ \Pi_{4}^\text{QCD, j} (Q^2) g_{\mu\nu} \mathbb{1} \nonumber \\
&+ \Pi_{5}^\text{QCD, j} (Q^2) p_{\mu} p_{\nu} \qslash 
+ \Pi_{6}^\text{QCD, j} (Q^2) q_{\mu} q_{\nu} \qslash
+ \Pi_{7}^\text{QCD, j} (Q^2) p_{\mu} q_{\nu} \qslash +\cdots\Big].
\end{align}
In Appendix~\ref{ap:DAs}, we present the structure $\Pi_{1}^\text{QCD, j} (Q^2)$ as an example in Eq.~\eqref{eq:exampleStr}. By equating the structures derived from both the physical and QCD sides, we can obtain the explicit analytical forms of the gravitational form factors for different flavors of the proton: $A^j(t)$, $J^j(t)$, $D^j(t)$, and $\bar{c}^j(t)$. In the following section, we provide the obtained gravitational form factors and their numerical analysis.

\section{Numerical results}\label{sec:numresults}

This section presents a detailed numerical analysis of the gravitational form factors of the proton for different flavors derived from the LCSR in the previous sections. To conduct this analysis, we utilize the distribution amplitudes of nucleons as derived from the work of Braun et al. (Ref.~\cite{Braun:2006hz}) at scale $\mu = 1$ GeV. These distribution amplitudes include hadronic parameters, as shown in Table~\ref{table:DAsparameters}. It is important to note that our calculations are performed in the chiral limit, with the mass of the proton set at  $m_N = 0.94$ GeV. Moreover, we use quantities such as the quark condensate $\langle \bar{q}q\rangle=(-0.24\pm 0.01)^3$~GeV$^3$ and the quark-gluon condensate parameter $m_0^2=0.8 \pm 0.1$~GeV$^2$~\cite{Ioffe:2005ym}, which are essential for determining the nucleon residue $\lambda_N$ as outlined in Eq.~\eqref{eq:residue}.

\begin{table}[!htb]
\centering	
\begin{tabular}{c*{12}{c}r}	
\hline
\hline
$f_N$ 
& \qquad $(5.0 \pm 0.5)\times 10^{-3} \text{GeV}^2$
& \qquad $f^{d}_1$ 
& \qquad
$0.40 \pm 0.05$ 
& \qquad $A^{u}_1$ 
& \qquad $0.38 \pm 0.15$
\\
$\lambda_1$ 
& \qquad $-(2.7 \pm 0.9)\times 10^{-2} \text{GeV}^2$ 
& \qquad $f^{d}_2$ 
& \qquad $0.22 \pm 0.05$
& \qquad $V^{d}_1$ 
& \qquad $0.23 \pm 0.03$
\\
$\lambda_2$ 
& \qquad $(5.4 \pm 1.9)\times 10^{-2} \text{GeV}^2$ 
& \qquad $f^{u}_1$ 
& \qquad $0.07 \pm 0.05$
\\
\hline
\end{tabular}
\caption{The parameters of the nucleon's distribution amplitudes.}
\label{table:DAsparameters}
\end{table}

In this method, three auxiliary parameters, $M^2, s_0$ and $\beta$,  determine the appropriate ranges, which directly affect the reliability of our results. First, we fix the Borel parameter $M^2$ within a range that ensures our results are approximately independent of this parameter. In the light-cone sum rules method, we utilize the distribution amplitudes (DAs) of the nucleon, which are derived for a fixed range of $1~\text{GeV}^2 \leqslant M^2 \leqslant 1.5~\text{GeV}^2$. The same applies to the second parameter $s_0$, which is the continuum threshold. This parameter is taken to be in the range $ 2.25 ~\text{GeV}^2 \leqslant s_0 \leqslant 2.40 ~\text{GeV}^2 $ based on the nucleon analyses. Finally, we consider the mixing parameter $\beta$ in the nucleon current, with $\beta = tan \theta$. While $\beta = -1$ corresponds to the well-known Ioffe current, we find its value to fall outside of the stable range that we desire. For the singlet case, where we do not consider flavors separately, the $\beta$ range is quite loose. The investigation of the up and down flavors, significantly narrows the range of arbitrary $\beta$, leading us to work with a fixed parameter, $\beta = -2 \pm 0.4$, which corresponds to $-0.53 \leqslant \cos\theta \leqslant -0.38$. We observe that up and down flavors make the physical quantities more sensitive to these auxiliary parameters compared to the $u+d$ analysis which finally contributes to the uncertainty of our findings.

\begin{figure}[!htb]
\centering
\includegraphics[scale=0.6]{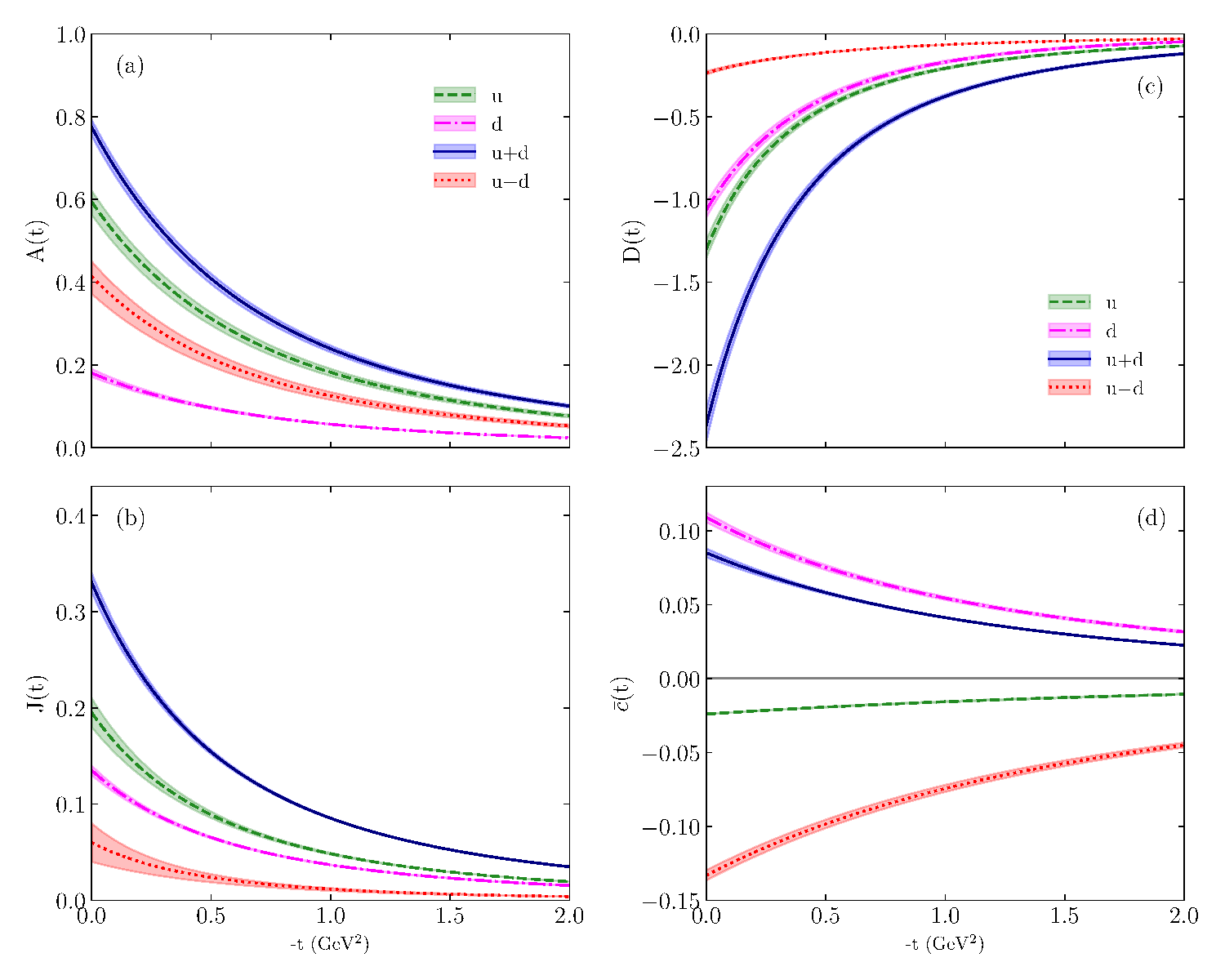}
\caption{\small 
The flavor-decomposed gravitational form factors of the proton are presented as a function of momentum transfer squared $(-t)$ at the scale $\mu = 1$ GeV. The curves represent different flavor contributions to the GFFs: the dashed green, dash-dotted pink, solid blue and dotted red curves indicate $u, d, u+d$ and $u-d$ flavor contributions, respectively. The shaded areas in the plot represent the uncertainty in the calculations. A p-pole fitting method is employed in this analysis.
}
\label{fig:GFFs}
\end{figure}

We derive numerical values for flavor-decomposed gravitational form factors of the proton within the range of $1 ~\text{GeV}^2 \leqslant -t~\leqslant 10 ~\text{GeV}^2$ using light-cone QCD sum rules. The results from the LCSR become singular in the vicinity of $t=0$. To obtain the distributions of these gravitational form factors across all momentum ranges, including $t=0$, we consider fitting the available data accordingly. The gravitational form factors of the proton for different flavors, as a function of the momentum transfer squared $-t$  are illustrated in Fig.~\ref{fig:GFFs}.
We utilize the p-pole fit, which is a well-established fitting method, commonly used in the literature for extrapolating gravitational form factors. The fit employed is as follows: 
\begin{equation}\label{eq:FitFun}
\mathcal{G}(t) = \frac{{\mathcal{G}}(0)}{\Big(1- g_{p}\,t\Big)^p},
\end{equation}
where ${\cal G} (0)$ represents the value of the form factors at $ t=0 $ (which is dimensionless), $p$ is also dimensionless, and $g_{p}$ has the dimension of inverse square energy (GeV$^{-2}$). Our obtained values for these parameters are provided in Table~\ref{table:fitparameters}. The contribution of the u-quark to the two form factors $A^{u+d}(t)$ and $J^{u+d}(t)$ is larger than that of the d-quark. This is expected, as the proton is composed of two u-quarks and one d-quark. However, the contributions from the u-quark and d-quark do not contribute equally to each form factor. Our findings indicate that all flavors of $D(t)$ form factor are negative and the absolute value of the u-quark contribution exceeds by a small amount that of the d-quark contribution.
The results for $D(t)$ form factor are consistent with an assumption about the values of $D$-term: $ D^u \approx D^d \approx D^q/2 $ as noted in~\cite{Polyakov:2018zvc}, where $q=u+d$ and D-term is the $D(t)$ form factor at zero momentum transfer. This condition suggests that the flavor singlet combination is dominant.
Our results are in accordance with this assumption, where the d-quark contributes almost equally to the u-quark. Our results for $\bar{c}(t)$ form factors show that the sign of u- and d-quark contributions are opposite and the absolute value of the down contributes more than that of up, also, the sign is positive for the singlet case of the EMT quark current, $\bar{c}^{u+d}$. The uncertainties in our resulting form factors stem from variations in the presented regions of $M^2$, $s_0$, and $\beta$, as well as inaccuracies in the input parameters and nucleon distribution amplitudes.

\begin{table}[!htb]
\centering	
\begin{tabular}{c*{3}{c} | c*{12}{c}}	
\hline
\hline
GFF & \qquad ${\cal G}(0)$  & \quad $g_{p}$~(GeV$^{-2}$) & 
\quad 
\hspace{0.6 cm} 
$p$ 
\hspace{1. cm}
& 
\quad
GFF & \qquad ${\cal G}(0)$  & \quad $g_{p}$~(GeV$^{-2}$) & \quad $p$
\\
\hline
$A^{u}(t)$ & \qquad $0.593(30)$  &\quad $0.436(2)$ &\quad $3.260(15)$ 
& 
\quad
$A^{u+d}(t)$ & \qquad $0.773(20)$  &\quad$0.429(8)$ &\quad $3.290(24)$  \\
$A^{d}(t)$ & \qquad $0.181(10)$ &\quad$0.405(31)$ &\quad$3.403(97)$
& 
\quad
$A^{u-d}(t)$ & \qquad $0.412(40)$  &\quad$0.449(8)$ &\quad $3.208(11)$  \\
$J^{u}(t)$ & \qquad $0.195(15)$  &\quad$0.678(63)$ &\quad $2.695(85)$
& 
\quad
$J^{u+d}(t)$ & \qquad $0.330(10)$  &\quad$0.658(11)$ &\quad $2.675(15)$  \\
$J^{d}(t)$ & \qquad $0.135(5)$  &\quad$0.625(63)$ &\quad $2.68(16)$
& 
\quad
$J^{u-d}(t)$ & \qquad $0.06(2)$  &\quad$0.71(29)$ &\quad $3.19(55)$  \\
$D^{u}(t)$ & \qquad $-1.300(55)$  &\quad$1.067(17)$ &\quad $2.520(17)$
& 
\quad
$D^{u+d}(t)$ & \qquad $-2.36(10)$  &\quad$0.808(39)$ &\quad $3.101(92)$  \\
$D^{d}(t)$ & \qquad $-1.064(45)$  &\quad$0.548(4)$ &\quad $4.188(44)$
& 
\quad
$D^{u-d}(t)$ & \qquad $-0.236(10)$  &\quad$0.892(80)$ &\quad $2.003(70)$ \\
$\bar{c}^{u}(t)$ & \qquad $-0.0240(2)$  &\quad$0.116(19)$ &\quad $3.96(49)$
& 
\quad
$\bar{c}^{u+d}(t)$ & \qquad $0.0850(30)$  &\quad$0.224(62)$ &\quad $3.74(85)$  \\
$\bar{c}^{d}(t)$ & \qquad $0.1090(32)$  &\quad$0.328(29)$ &\quad $2.47(16)$
& 
\quad
$\bar{c}^{u-d}(t)$ & \qquad $-0.1330(34)$  &\quad$0.172(3)$&\quad $3.68(2)$  \\
\hline
\end{tabular}
\caption{The p-pole fit parameters ${\cal G}(0)$, $g_{p}$ and $p$ of the flavor-decomposed GFFs in Fig.~\ref{fig:GFFs}. }
\label{table:fitparameters}
\end{table}



Our findings are shown alongside several recent studies of the flavor-decomposed GFFs at zero momentum transfer in Table~\ref{table:comp}. It is important to note that the scales used in these studies differ. All of these studies reported the quark contribution to GFFs; however, we focus exclusively on the valence quarks like Ref.~\cite{Won:2023cyd}, while other studies also included contributions from sea quarks, such as strange~\cite{Hackett:2023rif, Yao:2024ixu} and charm quarks~\cite{Yao:2024ixu}. The total contributions from quarks and gluons should satisfy $A^{q+g}(0)=1$, which is one of the well-known characteristics of the proton. 
By considering quark contributions, we get $A^{u+d}(0)=0.773(20)$, which is almost similar to the $A^q$ results from the symmetry-preserving CSM model when rescaled to match our scale, $\mu = 1$ GeV~\cite{Barone:2001sp, Hatta:2018sqd, Tanaka:2018nae}. The value of our resultant $A(0)$ for the up quark is approximately three times greater than that of the down quark, whereas other studies report a ratio of around two. The expected result for the  $J^{q+g}(0)$ is $1/2$, which is the spin of the proton (also a fundamental property of the particle). We drive $J^{u+d}(0)=0.330(10)$ for the singlet current. After rescaling the data from Table~\ref{table:comp} to align with our scale, we find that the lattice QCD derivation of $J^q(0)$ is relatively consistent with ours.
In this work, we find $J^u(0) / J^d(0) \approx 1.44 $, suggesting that the d-quark possesses a considerable contribution, similar to the findings in Ref.~\cite{Yao:2024ixu}.  
Our obtained $D^q(0)$ form factor is only close to the symmetry-preserving CSM model after rescaling to $\mu=1$ GeV.
We observe that the contributions of the up and down quarks are comparable in all the studies included in the table, indicating that both quarks contribute notably, also our results and CSM are the ones whose absolute values of up are greater than the down quark.

The $\bar{c}(t)$ form factor helps us understand the mechanics of the proton~\cite{Liu:2021gco,Polyakov:2018zvc}. Here, we obtain a positive value for the isoscalar, $\bar{c}^{u+d}(0) = 0.0850(30)$, and a negative value for the isovector case, $\bar{c}^{u-d}(0) = -0.1330(34)$. 
In contrast to the other form factors we have derived, the d-quark exhibits a significantly greater absolute value than the u-quark, along with the contributions from the single flavors' opposite signs.
The computed results of $\bar{c}(0)$ from the pion mean-field approach for the various flavors are presented in Table~\ref{table:comp}. The derivations for the u- and d-flavors from this calculation carry similar signs as ours.
In Ref.~\cite{Polyakov:2018exb}, the value $1.4 \times 10^{-2}$ is reported for the $\bar{c}^{u+d}(0)$ at $\mu^2\sim 0.4$ GeV$^2$. While the sign of this value aligns with our findings, it is not anywhere near our result when rescaled to our $\mu$. In Ref.~\cite{Hatta:2018sqd}, there is a disagreement regarding the sign of the $\bar{c}^{q}(0) \approx - 0.103$ (for $n_f =2$) in asymptotic limit ($\mu \rightarrow \infty$). In another study \cite{Liu:2021gco}, where the scale is $\mu = 2$ GeV, the reported value is $\bar{c}^{u+d}(0) = - 0.124(3)$. Generally, there is no consensus on the value of this form factor at zero momentum transfer, nor even on its sign. The only certain expectation is that the sum of quark and gluon contributions to this form factor should be zero. We expect that all calculations of $\bar{c}(t)$ yield a small value, regardless of its sign~\cite{Polyakov:2018exb, Polyakov:2018zvc}.

\begin{table}[!htb]
\centering	
\begin{tabular}{c*{12}{c}r}	
\hline
\hline
Model/approach & \qquad $A^{u}(0)$  & \quad $A^{d}(0)$ & \quad $A^{q}(0)$
& \quad $A^{u-d}(0)$ 
\\
\hline
This work ($\mu=1$ GeV) & \qquad $0.593(30)$  &\quad $0.181(10)$ &\quad $0.773(20)$ 
& \quad $0.412(40)$  \\
LQCD ($\mu=2$ GeV) & \qquad $0.3255(92)$ & \quad $0.1590(92)$ & \quad $0.510(25)$ & \quad -  
\\
symmetry-preserving of CSM ($\mu=2$ GeV)  & \qquad $0.328(15)$ & \quad $0.149(07)$ & \quad $0.584(13)$ & \quad -  
\\
pion mean-field ($\mu \approx 0.6$ GeV) & \qquad $0.66$ & \quad $0.34$ & \quad $1.00$ & \quad $0.32$ 
\\
\hline
 & \qquad $J^{u}(0)$  & \quad $J^{d}(0)$ & \quad $J^{q}(0)$
& \quad
$J^{u-d}(0)$
\\
\hline
This work & \qquad $0.195(15)$  &\quad $0.135(5)$ &\quad $0.330(10)$
& \quad $0.06(2)$  \\
LQCD & \qquad $0.2213(85)$ & \quad $0.0197(85)$ & \quad $0.251(21)$ & \quad - 
\\
symmetry-preserving & \qquad $0.164(08)$ & \quad $0.074(04)$ & \quad $0.292(06)$ & \quad -  
\\
pion mean-field  & \qquad $0.53$ & \quad $-0.03$ & \quad $0.50$ & \quad $0.56$ 
\\
\hline
 & \qquad $D^{u}(0)$  & \quad $D^{d}(0)$ & \quad $D^{q}(0)$
& \quad
$D^{u-d}(0)$
\\
\hline
This work & \qquad $-1.300(55)$  &\quad $-1.064(45)$ &\quad $-2.36(10)$
& \quad $-0.236(10)$  \\
LQCD & \qquad $-0.56(17)$ & \quad $-0.57(17)$ & \quad $-1.30(49)$ & \quad - 
\\
Symmetry-preserving  & \qquad $-1.019(49)$ & \quad $-0.463(22)$ & \quad $-1.820(43)$ & \quad - 
\\
pion mean-field  & \qquad $-1.12$ & \quad $-1.41$ & \quad $-2.53$ & \quad $0.29$ 
\\
\hline
 & \qquad $\bar{c}^{u}(0)$  & \quad $\bar{c}^{d}(0)$ & \quad $\bar{c}^{q}(0)$
& \quad
$\bar{c}^{u-d}(0)$
\\
\hline
This work & \qquad $-0.0240(2)$  &\quad $0.1090(32)$ &\quad $0.0850(30)$
& \quad $-0.1330(34)$  \\
LQCD & \qquad - & \quad - & \quad - & \quad -  
\\
Symmetry-preserving  & \qquad - & \quad - & \quad - & \quad - 
\\
pion mean-field  & \qquad $-0.04$ & \quad $0.04$ & \quad $0.00$ & \quad $-0.08$ 
\\
\hline
\end{tabular}
\caption{
Our flavor-decomposed gravitational form factors of the proton at $t = 0$ are presented alongside predictions from other models~\cite{Hackett:2023rif, Yao:2024ixu, Won:2023cyd}. In our work and the pion mean-field approach, the quark flavor combination is defined as $q = u + d$. In the lattice QCD study, it is extended to $q = u + d + s$, where $s$ represents the strange quark. In the symmetry-preserving of continuum Schwinger function method (CSM), $q = u_V + d_V + u_S + d_S + s + c$, where $V$ and $S$ denote valence and sea components of the light quarks, respectively, and $c$ represents the charm quark.
}
\label{table:comp}
\end{table}

\section{Mechanical properties}\label{sec:mech properties}


In this section, we study the mechanical properties of the proton and interpret the role of each flavor within its system. The $A(t)$ and $J(t)$ form factors
of proton are related to mass and spin, respectively. The $D(t)$ form factor determines the internal interactions, including the deformation and shape of the proton as well as its elastic properties. The renormalization-scale invariant quantity D-term is the $D(t)$ form factor at zero momentum transfer that does not get a fixed value by spacetime symmetries, unlike $A(0)$ and $J(0)$~\cite{Polyakov:1999gs}.
The internal pressure and shear forces are key components of proton mechanical structure that influence the spatial distribution of quarks and gluons. These forces are related to the $D(t)$ form factor which shows the significance of this form factor. These forces ensure stability, balance internal forces, and shape the proton. The stability conditions are derived from the elasticity theorem and the fact that the total EMT is conserved.

To facilitate our analysis, we work in the Breit frame with the following definitions for the kinematical variables $P^{\mu}$, $\Delta^{\mu}$ and momentum transfer squared $t$,
\begin{equation}\label{eq:Breit}
P^{\mu} = (E, \vec{0}), \qquad \Delta^{\mu} = (0, \boldsymbol{\Delta}),
\qquad t = \Delta^2 = - {\boldsymbol{\Delta}}^2 = 4 (m^2 - E^2).
\end{equation}
The different components of the static EMT include $T_{00}(r)$, which determines the energy density, and $T_{0k}(r)$, which describes the spatial distribution of momentum. In this section, we concentrate on the $T_{ij}(r)$ component, known as the stress tensor, as it is the central topic of our discussion. The stress tensor is decomposed in the following form,
\begin{equation}
T_{ij}(r)= (\frac{r^i r^j}{r^2}-\frac{1}{3}\delta_{ij}) s(r)+\delta_{ij}  p(r).
\label{eq:Tij}
\end{equation}
The differential equation below relates the pressure $p(r)$ and the shear force $s(r)$,
\begin{equation}
\label{eq:diff}
\frac{2}{3} s'(r) + \frac{2}{r} s(r) + p'(r) = 0,
\end{equation}
which originate from the conservation of the stress part $\nabla^i T_{ij}(r) = 0$ for the static EMT. A fundamental consequence of the EMT conservation is the von Laue condition, which states that the integral of the radial pressure over all space must vanish:
\begin{equation}
\label{eq:von Laue}
\int_{0}^{\infty} r^2 dr ~ p(r)= 0.
\end{equation}
This condition guarantees that internal forces balance, preventing any net expansion or contraction of the proton. This condition also holds for meta-stable and unstable systems~\cite{Perevalova:2016dln}. This is why von Laue is not sufficient for gaining a stable system.

In the Breit frame, the distributions of flavor-decomposed energy, pressure and shear force of the proton can be defined as combinations of the gravitational form factors as follows~\cite{Polyakov:2018zvc},
\begin{eqnarray}
&&\varepsilon^j(r) = m
\bigg[A^j(t) + \bar{c}^j(t) - \frac{t}{4 m^2} \Big(A^j(t) - 2 J^j(t) + D^j(t) \Big)      \bigg]_{FT},\label{eq:energyden}
\\
&&p^j_{0}(r) = \frac{1}{6m} \frac{1}{r^2} \frac{d}{dr} r^2 \frac{d}{dr} [D^j(t)]_{FT},\label{eq:p0}
\\
&&p^j_{1}(r) = \frac{1}{6m} \frac{1}{r^2} \frac{d}{dr} r^2 \frac{d}{dr} [D^j(t)]_{FT} - m [\bar{c}^j(t)]_{FT},\label{eq:p1}
\\
&&s^j(r) = -\frac{1}{4m} r \frac{d}{dr} \frac{1}{r} \frac{d}{dr} [D^j(t)]_{FT},\label{eq:s}
\end{eqnarray} 
where $j$ indicates the different flavors of quark and the three-dimensional Fourier transformation of GFFs is given by,
\begin{equation}
\label{eq:fourier}
[f(t)]_{FT} = \int \frac{d^3 \boldsymbol{\Delta}}{(2 \pi)^3} e^{-i \boldsymbol{\Delta}.\boldsymbol{r}} f(t)
= \frac{1}{4 \pi^2} \int_{-\infty}^{0} \frac{\sin[r\sqrt{-t}]}{r} f(t) dt,
\end{equation}
note that the last term is the simplified form of the Fourier transformation considering the proton to be spherically symmetric. The non-conserved form factor $\bar{c}^j(t)$ contributes to both energy density, Eq.~\eqref{eq:energyden}, and pressure, Eq.~\eqref{eq:p1}. We introduce two definitions of pressure: $p^j_{0}(r)$, which excludes $\bar{c}^j(t)$, and $p^j_{1}(r)$, which includes $\bar{c}^j(t)$. The exclusion of $\bar{c}^j(t)$ enables us to verify the validity of the proton's global and local stability criteria. The longitudinal force is defined in terms of the conserved pressure $p^j_{0}(r)$ as,
\begin{equation}
\label{eq:lforce}
F^{j}_{||}(r) = p^{j}_{0}(r) + \frac{2}{3} s^j(r).
\end{equation}
where it is the normal force per unit area $F_n(r) = F_{||}$. Here, we establish some local constrains, which unlike the von Laue condition, are not integrated over a range of radii. 
For local stability, it is necessary for the longitudinal force to be positive,
\begin{equation}
\label{eq:fn}
    F_{||}(r)>0,
\end{equation}
which means it is directed outwards, otherwise the proton would collapse. This stability condition leads to a negative D-term~\cite{Perevalova:2016dln, Burkert:2023wzr}. The shear force is typically positive in stable hydrostatic systems which means $s(r)>0$, that provides the second local stability condition.

\begin{figure}[!htb]
\includegraphics[scale=0.6]{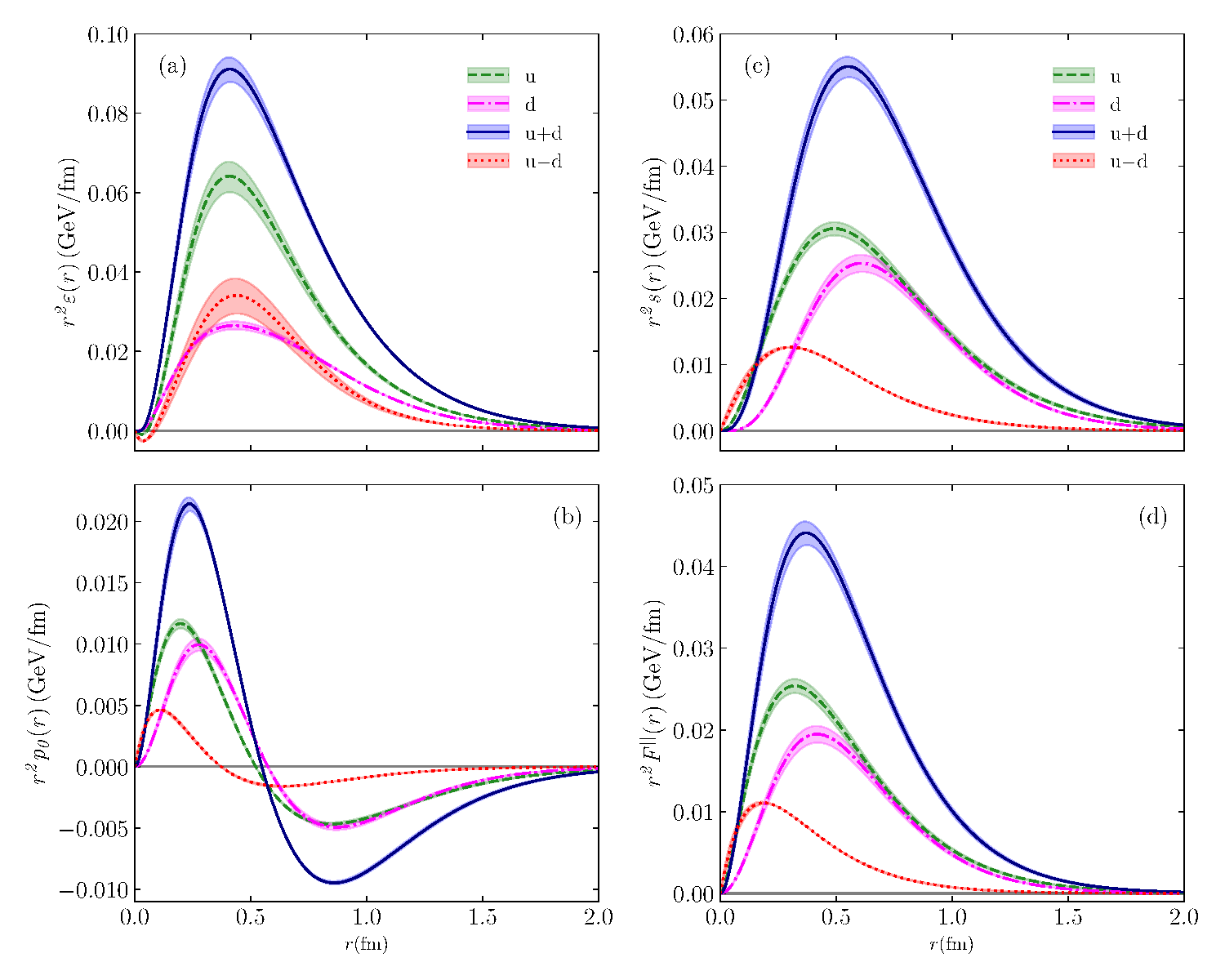}
\caption{\small
The proton spatial distributions of the energy density, pressure $p^j_{0}(r)$, shear force, and longitudinal force for each valence quark and their flavor combinations.
}
\label{fig:physdistribution}
\end{figure}

By decomposing the quark sector into its individual quarks and their combinations of valence flavors, we examine the impact of flavors in the internal proton system and whether these flavors meet the global and local stability conditions mentioned earlier. In Fig.~\ref{fig:physdistribution}, we present and compare the spatial distributions of the energy density, pressure $p^j_0(r)$, shear force, and longitudinal force for different flavors of proton, including their uncertainties. The isoscalar combination of $u+d$ displays the largest contribution in all the presented spatial distributions, with the u-quark generally contributing more significantly than the d-quark. However, this dominance of the u-quark in the pressure distribution does not hold across all ranges of $r$. In contrast, the isovector combination $u-d$ shows the smallest contribution in all spatial distributions, except for the energy density. The energy density peaks at approximately $r\approx0.4$ fm, which is almost the same for all flavor combinations. However, the peak of other distributions does not exhibit similar behavior for different flavors.

In Fig.~\ref{fig:physdistribution}$\,(a)$ the energy density of all flavor combinations is shown, as given by Eq.~\eqref{eq:energyden}. 
For a complete system (quarks$+$gluon), the total energy density is expressed as~\cite{Polyakov:2018zvc},
\begin{equation}
\label{eq:massNormal}
\varepsilon(r) = T_{00} (r) = m
\bigg[A(t) - \frac{t}{4 m^2} \Big(A(t) - 2 J(t) + D(t) \Big)      \bigg]_{FT}
\end{equation}
where the Fourier transformation is defined in Eq.~\eqref{eq:fourier}. Restricting our analysis to the quark sector of the EMT current, the non-conserved term $\bar{c}(t)$ emerges. As shown in Eq.~\eqref{eq:energyden}, the corresponding energy density for each flavor, $\varepsilon^j(r)$, includes the form factor $\bar{c}^j(t)$. For the complete system, the energy density is normalized as~\cite{Polyakov:2018zvc},
\begin{equation}
\label{eq:massNormal}
m = \int d^3 r ~\varepsilon(r).
\end{equation} 
Thus, integrating the total energy density over all space results in the total mass of the proton, $m$. By considering the contributions from different flavors, we can define a corresponding normalization condition of a complete system for each flavor, as follows,
\begin{equation}
\label{eq:mass}
m^j = \int d^3 r ~\varepsilon^j(r).
\end{equation}
where $m^j$ exclusively reflects the mass contributions arising from each flavor of the quark sector.
Using our energy distributions, we get $m^{u}/m^{u+d} \simeq 2/3$ and $m^{d}/m^{u+d} \simeq 1/3$, which are consistent with our expectations for the proton. This calculation of masses is performed naively, in a way that we assumed the proton always has the spherically symmetric form and neglecting the contributions from sea quarks and gluons, although this assumption is not entirely accurate~\cite{Miller:2003sa}.  This is only presented to compare the contributions of each quark. It is important to note that the calculations of mass are complex and do not fall within the scope of this paper. In Fig.~\ref{fig:physdistribution}$\,(b)$, we present the flavor decomposition of the pressure distribution as outlined in Eq.~\eqref{eq:p0}. Notably, we find that for each flavor, the pressure $p^j_{0}(r)$ satisfies the von Laue condition specified in Eq.~\eqref{eq:von Laue}. This implies that the positive pressure directed outwards is balanced by the negative pressure directed inwards, contributing to the stability of the proton. Our results align with other studies indicating that the inner region exhibits positive pressure while the outer region shows negative pressure.  As illustrated in Fig.~\ref{fig:physdistribution}$\,(b)$, the pressure for the isoscalar component changes sign at approximately $r\approx 0.55$ fm, which is roughly consistent with JLab data $r\approx 0.6$ fm~\cite{Burkert:2018bqq}. The positive pressure is more compact whereas the negative pressure is distributed over a larger range of $r$. Since the distributions of up and down quark are close to each other, we observe small values for the peak and valley of the isovector pressure.  In Fig.~\ref{fig:physdistribution}$\,(c)$, we present the shear force from our calculations, given by Eq.~\eqref{eq:s}.
The shear force corresponds to the traceless part of the stress tensor as described in Eq.~\eqref{eq:Tij}, making it insensitive to the non-conserving term of the EMT current. Consequently, the local condition $s(r) > 0$ should remain satisfied not only in the complete proton system but also within the quark sector. In Fig.~\ref{fig:physdistribution}$\,(c)$ we observe that this condition holds for the two individual quark flavors $u$, $d$, and for flavor combinations $u+d$ and $u-d$. Polyakov proposed a conjecture regarding the surface tension energy compared to the total energy of the  system in Ref.~\cite{Polyakov:2018zvc} where $\int d^3r s(r) \leqslant m$. We have examined this condition for our shear force and find that all cases satisfy the relation  $\int d^3r s^j(r) < m^j$. In Fig.~\ref{fig:physdistribution}$\,(d)$, we show the longitudinal force of the different flavors derived from Eq.~\eqref{eq:lforce}. As expected, the first local stability condition, as stated in Eq.~\eqref{eq:fn}, holds for the isoscalar case and remarkably for the individual flavors and isovector. It is important to note that we employed Eq.~\eqref{eq:lforce}, which is based on the $p_0^j(r)$ and is independent of the $\bar{c}(t)$ form factor. However, if we had specified the longitudinal force with respect to $p_1^j(r)$, we should have checked whether any modification to condition~\eqref{eq:fn} is needed due to the presence of $\bar{c}(t)$.

\begin{figure}[!htb]
\includegraphics[scale=0.6]{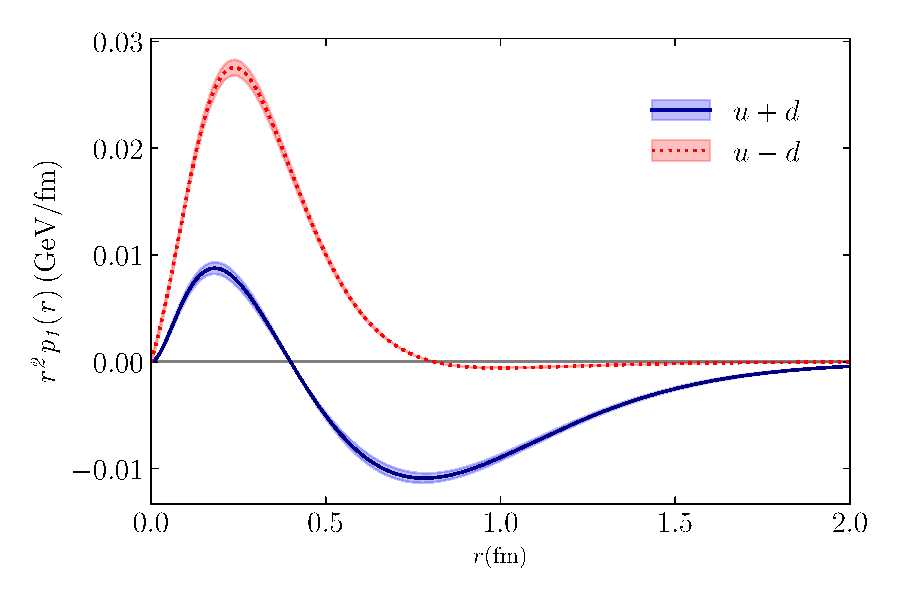}
\caption{\small 
The distributions of pressure $p^j_{1}(r)$ for the isoscalar and isovector proton flavor combinations.
}
\label{fig:noncp}
\end{figure}

In the quark sector of the EMT current, the non-conserved $\bar{c}(t)$ appears in the definition of pressure which is presented in Eq.~\eqref{eq:p1} as $p^j_{1}(r)$. Therefore it does not satisfy the von Laue condition, highlighting the role of $\bar{c}(t)$ in proton stability. When gluon contributions are included, the $\bar{c}(t)$ form factor vanishes, stating the relation $\sum_{q} \bar{c}^q(t) + \bar{c}^g(t) = 0$ which was shown in the $p_0^j(r)$ pressure in Eq.~\eqref{eq:p0}. A vanishing $\bar{c}(t)$ form factor indicates an internal balance between repulsive and attractive pressures within the proton i.e. the von Laue condition. In Fig.~\ref{fig:noncp}, we present the spatial distribution of the form factor $p^j_1(r)$ for isoscalar and isovector flavor contributions, as derived from Eq.~\eqref{eq:p1}. It is observed that the non-conserved $\bar{c}(t)$ form factor results in a slight deviation from the von Laue condition in the isoscalar case. This deviation is more intense in the isovector case, where the integration of pressure over the entire space yields a significantly large positive value. This indicates that the gluon plays a more crucial role when the proton is in the isovector state. Figure.~\ref{fig:noncp} highlights the necessity of gluon contributions in maintaining a stable proton system.

\begin{figure}[t]
\includegraphics[scale=0.43]{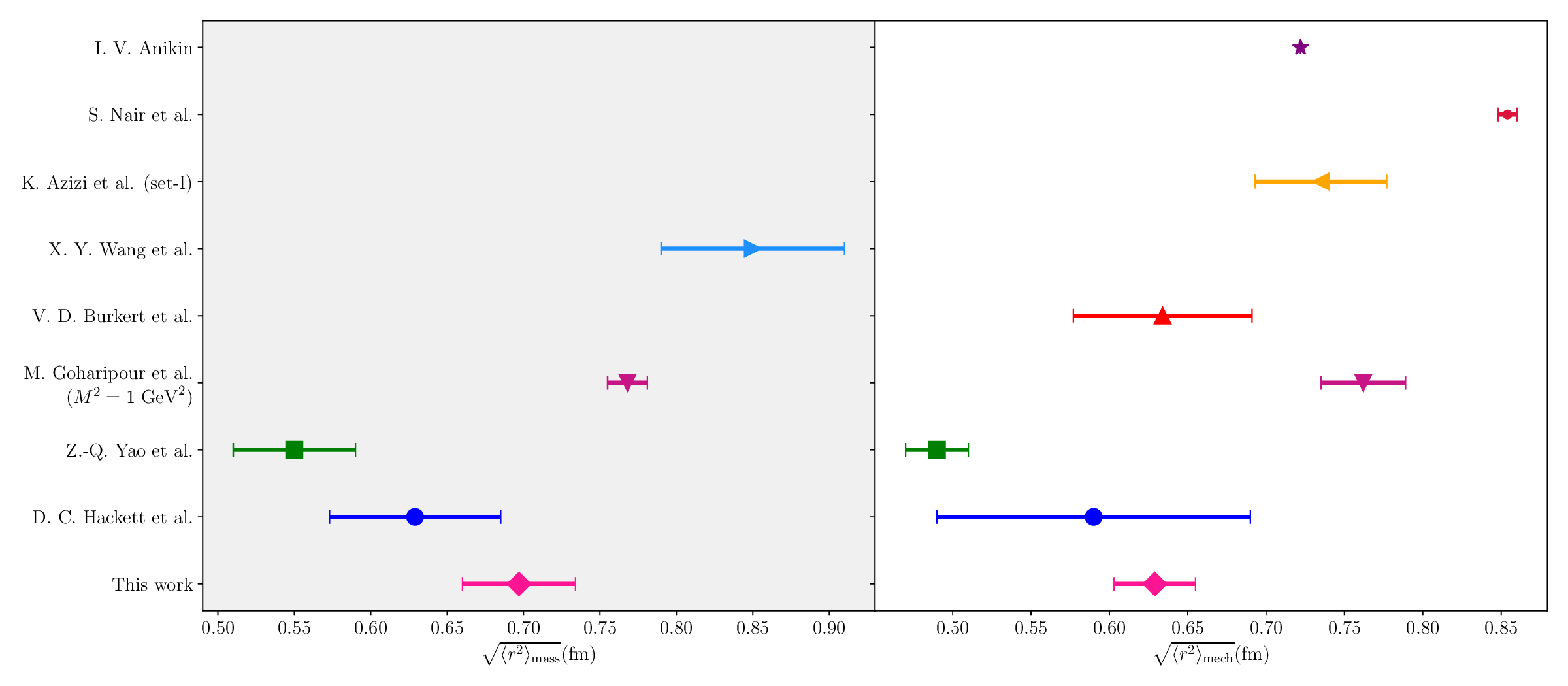}
\caption{\small Comparison of mass and mechanical radii results for the EMT quark current from our light-cone QCD calculations and other approaches~\cite{Hackett:2023rif, Yao:2024ixu, Goharipour:2025lep, Burkert:2023atx, Wang:2022zwz, Azizi:2019ytx, Nair:2024fit, Anikin:2019kwi} is depicted. The right hand side of the plot compares the mechanical and the left hand side the mass radii. The horizontal axes is in fermi unit.}
\label{fig:DistriDistortion}
\end{figure}

The mass radius of the different flavors of proton is defined as,
\begin{equation}
\label{eq:massR}
\langle r^2_j\rangle_{\text{mass}} = 
\frac{\int d^3 \boldsymbol{r} r^2 \varepsilon^j(r)}{\int d^3 \boldsymbol{r} \varepsilon^j(r)}.
\end{equation}
The mass radius of the proton describes the spatial distribution of its mass or to be more exact its energy. We consider the quark sector in this context, consequently the definition of energy density includes the term $\bar{c}(t)$ in its derivation (see Eq.~\eqref{eq:energyden}). Also the mechanical radius of the flavor-decomposed proton is as follows,
\begin{equation}
\label{eq:mechR}
\langle r^2_j\rangle_{\text{mech}} = 
\frac{\int d^3 \boldsymbol{r} r^2 F^{j}_{||}(r)}{\int d^3 \boldsymbol{r} F^{j}_{||}(r)}.
\end{equation}
The mechanical radius has been obtained using $F^j_{||}(r)$, which is partly dependent on $p^j_0(r)$. The definition of the mechanical radius is based on the conservation of the energy-momentum tensor current, and it seems illogical to use a pressure derived from a non-conserved EMT, namely $p^j_1(r)$, to calculate it. We obtain the mass and mechanical radii of the proton from the singlet EMT quark current as follows,
\begin{eqnarray}
\label{eq:OurRadii}
&&\sqrt{\langle r^2_p\rangle_{\text{mass}}} = 
0.697(37) ~\text{fm},
\\
&&\sqrt{\langle r^2_p\rangle_{\text{mech}}} = 
0.629(26) ~\text{fm}.
\end{eqnarray} 
Both the resultant mass and mechanical radii are smaller than the charge radius as expected. Additionally, the mechanical radius is smaller than the mass radius. This behavior agrees with some methods~\cite{Hackett:2023rif, Yao:2024ixu, Goharipour:2025lep} and shows that our obtained energy of the system is saturated in a larger volume than the internal forces, specifically the longitudinal force. A comparison of our mass and mechanical radii for the quark contribution with other recent theoretical and experimental calculations is shown in Fig.~\ref{fig:DistriDistortion}. The resulting mass radius of the quark sector aligns with lattice results from Ref.~\cite{Hackett:2023rif}, although it does not coincide with findings from other sources, such as~\cite{Yao:2024ixu}. Additionally, the quark mechanical radius falls within the range of the lattice and DVCS findings~\cite{Hackett:2023rif,Burkert:2023atx}. However, this radius does not overlap with results from Refs.~\cite{Nair:2024fit, Yao:2024ixu}. Furthermore, our results are not consistent with previous findings of LCSR method~\cite{Anikin:2019kwi,Azizi:2019ytx}. 
Our result for the mass radius is close to the GPD result, whereas our mechanical radius is inconsistent with GPD finding~\cite{Goharipour:2025lep}.

In Table.~\ref{table:radiiFlavor}, we provide the mass and mechanical radii for various flavors.
\begin{table}[b]
\centering	
\begin{tabular}{c*{12}{c}r}	
\hline
\hline
\rule{0pt}{3.5ex}
PF & \qquad $\sqrt{\langle r^2_j\rangle_{\text{mass}}}$ (fm)  & \quad $\sqrt{\langle r^2_j\rangle_{\text{mech}}}$ (fm) & \quad PF
& \qquad
$\sqrt{\langle r^2_j\rangle_{\text{mass}}}$ (fm) & \qquad $\sqrt{\langle r^2_j\rangle_{\text{mech}}}$ 
\rule[-2.5ex]{0pt}{0pt}  
\\
\hline
$u$ & \qquad $0.689(40)$  &\quad $0.615 (27)$  
& \qquad
$u + d$ & \qquad $0.697(37)$  &\quad$0.629(26)$  \\
$d$ & \qquad $0.730 (43)$ &\quad$0.639(28)$ 
& \qquad
$u - d$ & \qquad $0.644 (43)$  &\quad$0.456(22)$   \\
\hline
\end{tabular}
\caption{Mass and mechanical radii of singlet and triplet flavors in proton. PF is the shortened form for proton's flavors. }
\label{table:radiiFlavor}
\end{table}
We notice the derived radii for singlet and triplet currents are within the same order. Notably, we observe the mechanical radius is smaller than the mass radius for each flavor. The confinement phenomenon restricts us from discussing radii associated with single flavors. In this study, we compute these radii to see what is their effective range of energy and longitudinal force in proton system. In examining the energy and longitudinal distributions of the isovector, $u-d$, we notice a more rapid decline relative to the other quark flavors. This is probably the cause behind the small values of the mechanical and mass radii of isovector case compared to the other flavors. Despite the sensitivity of the subject, if we allow ourselves some freedom of thought, we might conclude that the up quarks construct a diquark subsystem alongside the down quark within the proton. Based on the derived mass and mechanical radii for both up and down quarks, it appears that the down quark exhibits greater freedom than the up quark.

\section{Conclusion }\label{sec:conclusion}

In this work, we investigated the flavor-decomposed gravitational form factors (GFFs) of the proton and their mechanical characteristics using the light-cone sum rules (LCSR) framework. Our calculations were performed using nucleon distribution amplitudes (DAs) up to twist-6 accuracy at an energy scale of $\mu = 1$ GeV. By incorporating all forms of the Borel transformations applicable to our calculations and analyzing key mechanical properties—including energy, pressure, shear force, and longitudinal force for different quark flavors—we provided a more thorough,  complete  and accurate investigation compared to the  previous studies.  Our findings reveal that for the singlet and triplet quark currents, the form factors 
$A(t)$ and $J(t)$ remain positive. Remarkably, the up-quark is dominant over the down-quark in the proton's mass and spin form factors. Additionally, we observe that the $D(t)$ form factor is negative across all flavors, with the absolute value of the u-quark exceeding that of the d-quark. In contrast to the conserved form factors, the non-conserved $\bar{c}(t)$ form factor exhibits a significantly larger contribution for d-quark than that of the u-quark, with the contributions from up and down quarks having opposite signs.

The gravitational form factors eventually build up the energy, pressure, and shear force distributions within the proton. Similar to the conserved GFFs, the u-quark contributes more significantly than the d-quark to the energy, shear force, and longitudinal force distributions. This study presents the pressure distribution in two forms: one that disregards the contribution of the $\bar{c}(t)$ form factor, denoted as $p_0(r)$, and the other considers its effects, referred to as $p_1(r)$. The first definition of pressure, and its application in describing the longitudinal force, satisfies both global and local stability conditions across all flavors. However, the pressure $p_1(r)$ fails to meet the primary stability criterion, the von Laue condition, which highlights the necessity of gluons in retaining a stable proton system: $\sum_q\bar{c}^q + \bar{c}^g = 0$. The $\bar{c}(t)$ form factor serves as an indicator of the quarks' role in the proton’s mechanical structure and quantifies the interaction between the quark and gluon subsystems. Notably, since the shear force is insensitive to the trace part of the energy-momentum tensor, it remains positive regardless of considering the whole  system or a part of the proton.

The mass and mechanical radii were extracted from the energy and longitudinal force distributions for the singlet case, yielding 
$\langle r^2_p\rangle^{1/2}_{\text{mass}} = 
0.697(37) ~\text{fm}$ 
and
$\langle r^2_p\rangle^{1/2}_{\text{mech}} = 
0.629(26) ~\text{fm}$ 
respectively. These values demonstrate good agreement with theoretical predictions and empirical measurements. Furthermore, they are smaller than the proton charge radius, consistent with the underlying nature of electromagnetic and strong interactions. Additionally, the mass and mechanical radii are presented for other flavors. We apprehend the mechanical radius is less than the mass radius for all the derived radii.

\begin{acknowledgments}

We thank Dimitra Pefkou for providing us with their Lattice QCD results. Z. Dehghan and K. Azizi are thankful to the Iran National Science Foundation (INSF) for the financial support provided for this research under project number 4025645.

\end{acknowledgments}

\appendix
\section{Some supplementary materials} \label{ap:DAs}

\setcounter{equation}{0}
\renewcommand{\theequation}{\Alph{section}.\arabic{equation}}

In this appendix, we present specific details of the QCD side of the Light-Cone Sum Rule (LCSR) correlation function. We explore the expansion of the nucleon matrix element of the three-quark operator in terms of nucleon distribution amplitudes (DAs), as discussed in the literature~\cite{Braun:2006hz},
\begin{eqnarray}
&&\langle 0|\epsilon^{abc} u_{\sigma}^a(a_1 x) u_{\theta}^b(a_2 x) d_{\phi}^c(a_3 x)|N(p)\rangle 
\nonumber \\
&&= \frac{1}{4}\Big\{ {\cal S}_1 m_N C_{\sigma \theta} \left(\gamma_5 N\right)_\phi +{\cal S}_2 m_N^2 C_{\sigma \theta} \left(\!\not\!{x} \gamma_5 N\right)_\phi 
+ {\cal P}_1 m_N
\left(\gamma_5 C\right)_{\sigma \theta} N_\phi + {\cal P}_2 m_N^2 \left(\gamma_5 C \right)_{\sigma \theta} \left(\!\not\!{x} N\right)_\phi
\nonumber \\
& &+ \Big({\cal V}_1 +\frac{m_N^2 x^2 }{4}{\cal V}_1^M \Big)  \left(\!\not\!{P}C \right)_{\sigma \theta} \left(\gamma_5 N\right)_\phi + {\cal V}_2 m_N \left(\!\not\!{P} C \right)_{\sigma \theta}
\left(\!\not\!{x} \gamma_5 N\right)_\phi  + {\cal V}_3 m_N  \left(\gamma_\mu C \right)_{\sigma \theta}\left(\gamma^{\mu} \gamma_5 N\right)_\phi
\nonumber \\
&& + {\cal V}_4 m_N^2 \left(\!\not\!{x}C \right)_{\sigma \theta} \left(\gamma_5 N\right)_\phi+ {\cal V}_5 m_N^2 \left(\gamma_\mu C \right)_{\sigma \theta} \left(i
\sigma^{\mu\nu} x_\nu \gamma_5 N\right)_\phi + {\cal V}_6 m_N^3 \left(\!\not\!{x} C \right)_{\sigma \theta} \left(\!\not\!{x} \gamma_5 N\right)_\phi
\nonumber \\
&& + \Big({\cal A}_1 +\frac{m_N^2 x^2 }{4}{\cal A}_1^M \Big)   \left(\!\not\!{P}\gamma_5 C \right)_{\sigma \theta} N_\phi + {\cal A}_2 m_N \left(\!\not\!{P}\gamma_5 C \right)_{\sigma \theta} \left(\!\not\!{x}
N\right)_\phi  + {\cal A}_3 m_N \left(\gamma_\mu \gamma_5 C \right)_{\sigma \theta}\left( \gamma^{\mu} N\right)_\phi
\nonumber \\
&& + {\cal A}_4 m_N^2 \left(\!\not\!{x} \gamma_5 C \right)_{\sigma \theta} N_\phi + {\cal A}_5 m_N^2 \left(\gamma_\mu \gamma_5 C \right)_{\sigma \theta} \left(i
\sigma^{\mu\nu} x_\nu N\right)_\phi + {\cal A}_6 m_N^3 \left(\!\not\!{x} \gamma_5 C \right)_{\sigma \theta} \left(\!\not\!{x} N\right)_\phi
\nonumber \\
&& + \Big({\cal T}_1 +\frac{m_N^2 x^2 }{4}{\cal T}_1^M \Big)  \left(P^\nu i \sigma_{\mu\nu} C\right)_{\sigma \theta} \left(\gamma^\mu\gamma_5 N\right)_\phi + {\cal T}_2 m_N \left(x^\mu P^\nu i \sigma_{\mu\nu}
C\right)_{\sigma \theta} \left(\gamma_5 N\right)_\phi \nonumber \\ 
&&+ {\cal T}_3 M \left(\sigma_{\mu\nu} C\right)_{\sigma \theta}
\left(\sigma^{\mu\nu}\gamma_5 N \right)_\phi+{\cal T}_4 m_N \left(P^\nu \sigma_{\mu\nu} C\right)_{\sigma \theta} \left(\sigma^{\mu\varrho} x_\varrho
\gamma_5 N\right)_\phi \nonumber
\\&& + {\cal T}_5 m_N^2 \left(x^\nu i \sigma_{\mu\nu} C\right)_{\sigma \theta} \left(\gamma^\mu\gamma_5 N \right)_\phi 
+ {\cal T}_6 m_N^2 \left(x^\mu P^\nu i \sigma_{\mu\nu} C\right)_{\sigma \theta} \left(\!\not\!{x} \gamma_5
N \right)_\phi \nonumber
\\&&+ {\cal T}_7 m_N^2 \left(\sigma_{\mu\nu} C\right)_{\sigma \theta}
\left(\sigma^{\mu\nu} \!\not\!{x} \gamma_5 N \right)_\phi + {\cal T}_8 m_N^3 \left(x^\nu \sigma_{\mu\nu} C\right)_{\sigma \theta} \left(\sigma^{\mu\varrho} x_\varrho
\gamma_5 N \right)_\phi \Big\}\,
\label{eq:das-def}
\end{eqnarray}
The ``calligraphic" DAs in the previous equation can subsequently be expressed in terms of ``direct" DAs, up to twist-6 accuracy, as delineated below,
\begin{align}
\mathcal{S}_1 =& S_1,                       \hspace{4.3cm} \mathcal{S}_2=\frac{1}{2px}\big(S_1-S_2\big) ,\nonumber\\
\mathcal{P}_1=&P_1,                        \hspace{4.3cm}\mathcal{P}_2=\frac{1}{2px}\big(P_2-P_1\big)
\nonumber\\
\mathcal{V}_1=&V_1,                       \hspace{4.3cm} \mathcal{V}_2=\frac{1}{2px}\big(V_1-V_2-V_3\big), \nonumber\\
\mathcal{V}_3=&V_3/2,                    \hspace{4cm} \mathcal{V}_4=\frac{1}{4px}\big(-2V_1+V_3+V_4+2V_5\big),\nonumber\\
\mathcal{V}_5=&\frac{1}{4px} \big(V_4-V_3 \big) ,     \hspace{2.62cm}\mathcal{V}_6=\frac{1}{4(px)^2}\big(-V_1+V_2+V_3+V_4 + V_5-V_6\big),
\nonumber\\
\mathcal{A}_1=&A_1,                      \hspace{4.3cm} \mathcal{A}_2=\frac{1}{2px}\big(-A_1+A_2-A_3\big),\nonumber\\
\mathcal{A}_3=&A_3/2,                              \hspace{4.cm}\mathcal{A}_4=\frac{1}{4px}\big(-2A_1-A_3-A_4+2A_5\big), \nonumber\\
\mathcal{A}_5=&\frac{1}{4px}\big(A_3-A_4\big),                \hspace{2.5cm}\mathcal{A}_6=\frac{1}{4(px)^2}\big(A_1-A_2+A_3+A_4-A_5+A_6\big)
\nonumber\\
\mathcal{T}_1=&T_1,                                  \hspace{4.5cm} \mathcal{T}_2=\frac{1}{2px}\big(T_1+T_2-2T_3\big), \nonumber\\
\mathcal{T}_3=&T_7/2,                                    \hspace{4.2 cm} \mathcal{T}_4=\frac{1}{2px}\big(T_1-T_2-2T_7\big),\nonumber\\
\mathcal{T}_5=&\frac{1}{2px}\big(-T_1+T_5+2T_8\big),                        \hspace{1.32cm}\mathcal{T}_6=\frac{1}{4(px)^2}\big(2T_2-2T_3-2T_4+2T_5+2T_7+2T_8\big),
\nonumber\\  \mathcal{T}_7=&\frac{1}{4px}\big(T_7-T_8\big),                 \hspace{2.62cm} \mathcal{T}_8=\frac{1}{4(px)^2}\big(-T_1+T_2 +T_5-T_6+2T_7+2T_8\big).
\label{eq:DAsdirect}
\end{align}
The direct distribution amplitudes (DAs), $F = S_i, P_i, V_i, A_i, T_i$ are defined in terms of $a_i p x$ and is expressed as follows,
\begin{equation}
 F(a_i px) = \int dx_1 dx_2 dx_3 ~ \delta {(x_1 + x_2 + x_3 -1)}~
   e^{-ipx \sum_i {x_i a_i}}~ F(x_i),
\label{eq:integ}   
\end{equation}
with $i=1, 2, 3$ and $0 < x_i < 1$ such that $\sum_i x_i = 1$. The $x_i$ values represent the longitudinal momentum fractions carried by the quarks within the nucleon. The comprehensive representations of the nucleon distribution amplitudes are presented in Ref.~\cite{Braun:2006hz}. 

Since the complete expression of each term of QCD correlation function is too long to be included here, we present a segment of the first structure in Eq.~\eqref{eq:QCDStr} of the specific flavor j as follows,
\begin{align}
\Pi_{1}^\text{QCD, j} (Q^2) &= 
\frac{m^3}{M^2} (1 + \beta) \int_{x_0}^{1} d\eta \int_{\eta}^{1} d\xi \int_{\xi}^{1} d\kappa \int_{\kappa}^{1} dx_3 \int_{0}^{1-x_3} dx_1 \frac{T_1[x_1, 1- x_1 - x_3, x_3]}{\eta}
e^{- s(\eta,Q^2)/M^2}
\nonumber\\
&+ \frac{m^3}{2 M^2} (1 - \beta) \int_{x_0}^{1} d\xi \int_{\xi}^{1} d\kappa \int_{\kappa}^{1} dx_3 \int_{0}^{1-x_3} dx_1 {A_1[x_1, 1- x_1 - x_3, x_3]}
e^{- s(\xi,Q^2)/M^2}
\nonumber\\
&- \frac{m^3}{2 M^2} (1 - \beta) \int_{x_0}^{1} d\kappa \int_{\kappa}^{1} dx_3 \int_{0}^{1-x_3} dx_1 {A_2[x_1, 1- x_1 - x_3, x_3]}
e^{- s(\kappa,Q^2)/M^2}
\nonumber\\
&+ \frac{m}{2} (1 - \beta) \int_{x_0}^{1} dx_3 \int_{0}^{1-x_3} dx_1  \;  x_3 A_3[x_1, 1 - x_1 - x_3, x_3]
e^{- s(x_3,Q^2)/M^2}
\nonumber\\
&- \frac{m^3}{2} (1 + \beta) \int_{x_0}^{1} dx_3 \int_{0}^{1-x_3} dx_1  \;  \frac{x_0^3 \; P_2[x_1, 1 - x_1 - x_3, x_3]}{Q^2+m^2 x_0^2}
e^{- s_0/M^2} + \cdots,
\label{eq:exampleStr}
\end{align}


\end{document}